\documentstyle[12pt]{article}

\makeatletter
\def\section{\@startsection{section}{1}{\z@}{-3.5ex plus -1ex minus -.2ex}
{2.3ex plus .2ex}{\large\bf}}
\makeatother
 
\makeatletter
\@addtoreset{equation}{section}

\makeatother

\def\la{\langle}
\def\ra{\rangle}

\def\lb{\lbrack}
\def\rb{\rbrack}

 \def\Slash#1{
  \begin{picture}(5,6)(0,0)
  \put(-.7,-1.2){\line(5,6)6}
  \end{picture}
  \kern-.8em#1}
 \def\slash#1{
  \begin{picture}(5,6)(0,0)
  \put(-1.5,-1.7){\line(5,6)5}
  \end{picture}
  \kern-.8em#1}

\def\tr{\mbox{tr}}
\def\e{\epsilon}
\def\g5{\gamma_5}
\def\hg5{\hat{\gamma}_5}

\def\tlambda{\widetilde{\lambda}}

\def\Qlatmr1{Q_{lat}^{(m=r=1)}}

\def\be{\begin{eqnarray}}
\def\ee{\end{eqnarray}}

\def\tchi{\widetilde{\chi}}
\def\T{{\cal T}}
\def\tpsi{\widetilde{\psi}}

\parskip=\medskipamount

\begin{document}

\input epsf

\vspace{8mm}

\begin{center}
 
{\Large \bf General bounds on the Wilson-Dirac operator}
\\

\vspace{0.3ex}

\vspace{12mm}

{\large David H. Adams}

\vspace{4mm}

Instituut-Lorentz for Theoretical Physics, Leiden University,\\ 
Niels Bohrweg 2, NL-2333 CA Leiden, The Netherlands \\

\vspace{1ex}

email: adams@lorentz.leidenuniv.nl

\end{center}

\begin{abstract}
Lower bounds on the magnitude of the spectrum of the Hermitian Wilson-Dirac operator
$H(m)$ have previously been derived for $0<m<2$ when the lattice gauge field satisfies 
a certain smoothness condition. In this paper lower bounds are derived for 
$2p-2<m<2p$ for general $p=1,2,\dots,d$ where $d$ is the spacetime
dimension. The bounds can alternatively be viewed as localisation bounds on the real 
spectrum of the usual Wilson-Dirac operator.
They are needed for the rigorous evaluation of the classical continuum limit of the
axial anomaly and index of the overlap Dirac operator at general values of $m$, and
provide information on the topological phase structure of overlap fermions. 
They are also useful for understanding the instanton size-dependence of the real
spectrum of the Wilson-Dirac operator in an instanton background.

\end{abstract}

\section{Introduction}

It is well-known from numerical studies (see, e.g., \cite{Itoh,Gatt}) that in smooth 
gauge backgrounds in $d$ dimensions the real eigenvalues of the Wilson-Dirac operator
are localised around the values $0,2,4,\dots,2d$ (in units of inverse lattice 
spacing, and with Wilson parameter $r=1$). In this paper we give an analytic derivation
of this numerical observation. Our smoothness condition is the ``admissibility
condition'' of \cite{L(local),Neu(bound)}:
\be
||1-U(p)||\;\le\;\e\qquad\quad\forall\,\mbox{plaquette}\ p.
\label{1.1}
\ee
Since the plaquette variable has the expansion $U(p)=1-a^2F_{\mu\nu}(x)+O(a^3)$
in powers of lattice spacing $a$, (\ref{1.1}) can be regarded as an approximate 
smoothness requirement on the curvature of the lattice gauge field. If $U$ is 
the lattice transcript of a smooth continuum gauge field then (\ref{1.1}) is
automatically satisfied for any $\e>0$ when the lattice is sufficiently fine.

In fermionic definitions of the topological charge of lattice gauge fields the 
low-lying real eigenmodes of the Wilson-Dirac operator $D_w$ are interpreted as
would-be zero-modes, while the other real eigenmodes are interpreted as would-be
doubler modes. This interpretation relies on the real eigenvalues being
localised as described above, which is not the case in general for arbitrary rough
gauge fields. The localisation result on the real spectrum of $D_w$
derived in this paper provides a specific analytic criterion under which the
localisation is guaranteed. It is also of interest in connection with the overlap
fermion formulation on the lattice \cite{ov,Neu}. This is because a real eigenmode 
for the Wilson-Dirac operator is equivalent to a zero-mode for the Hermitian 
Wilson-Dirac operator with negative mass parameter:
\be
D_w\psi=\frac{m}{a}\psi\qquad\Leftrightarrow\qquad{}H(m)\psi\,\equiv\,\g5(aD_w-m)\psi=0
\label{1.2}
\ee
Localisation of real eigenvalues of $D_w$ around $0,2,\dots,2d$ (in units of $1/a$)
is therefore equivalent to absence of zero-modes for $H(m)$, i.e. to existence of
non-zero lower bounds on $|H(m)|$, when $m$ is away from these values.
This implies a topological phase structure for the overlap Dirac operator \cite{Neu}
$D_{ov}=\frac{1}{a}(1+\g5{}H(m)/|H(m)|)$, since the index of $D_{ov}$ (a well-defined 
integer) is locally independent of $m$ but can jump at the values 
for which $H(m)$ has zero-mode(s). 
The topological phase structure of $D_{ov}$ has previously been studied 
in ref.'s \cite{Chiu,Sch}. The bounds derived in this
paper lead to analytic information on the topological phases which complements
the numerical results of those papers.

Furthermore, a non-zero lower bound on $|H(m)|$ allows the locality of the overlap
Dirac operator, and its smooth dependence gauge field, to be analytically
established \cite{L(local)} (see also \cite{Neu(rant)}).
The general bounds derived in this paper allow the unnatural
restriction $0<m<2$ on the results of \cite{L(local)} to be removed.
These bounds are also required for the rigorous evaluation of the classical
continuum limits of the axial anomaly and index of the overlap Dirac operator 
\cite{DA(AP),DA(JMP)}.\footnote{Other evaluations of the classical continuum limit
of the axial anomaly (less rigorous, and not using a lower bound on $|H(m)|$)
have been given in \cite{Kiku,Fuji,Suz,Nagao}.} As a final application we will discuss
qualitative implications of the bounds for the instanton size-dependence
of the real spectrum of the Wilson-Dirac operator in an instanton background.

The paper is organised as follows. In Section 2 the previously derived lower bounds 
on $|H(m)|$ are summarised and the new general bounds are formulated. The new bounds 
are derived in Section 3. The derivation is rather technical and not so illuminating,
so in Section 4 we supplement it with with a heuristic argument which provides a 
clearer intuitive understanding of why the bounds exist. The heuristic considerations
are further developed to give an analytic explanation of properties of the spectral
flow of $H(m)$ previously observed in numerical studies. 
In Section 5 the above-mentioned applications of the bounds are discussed, and the
results of the paper are summarised in Section 6.
A generalisation of the bounds from the standard case of
Wilson parameter $r=1$ to general values $r>0$ is given in an appendix.

\section{Summary of previous bounds and formulation of the new bounds}

For $m\le0$ and $m\ge{}2d$ ($d$=spacetime dimension) it is well-known that
$|H(m)|\ge|m|$ and $|H(m)|\ge{}m\!-\!2d$, respectively; see, e.g., 
\cite{ov,Edwards,Gatt}. (We review these bounds and generalise them to 
arbitrary values of the Wilson parameter $r$ in the Appendix.) By (\ref{1.2}) this
implies that the real eigenvalues of $D_w$ (in units of $1/a$) must lie in the 
interval $[0,2d]$. 

In \cite{L(local),Neu(bound)} lower bounds of the form
\be
|H(m)|\;\ge\;\sqrt{1-c_1\,\e}-|1-m|
\label{2.1}
\ee
were derived when the lattice gauge field satisfies the smoothness condition 
(\ref{1.1}). The currently sharpest bound has $c_1=6(2+\sqrt{2})\approx20.5$ in 
4 dimensions \cite{Neu(bound)} and generalises to
$c_1=(2+\sqrt{2})d(d-1)/2$ in $d$ dimensions. Clearly (\ref{2.1}) can only be a 
nontrivial lower bound if $\e<1/c_1$ and $|1-m|<\sqrt{1-c_1\e}$. The latter implies
$0<m<2$. Lower bounds on $|H(m)|$ in the ``doubler regions'' 
$2<m<4$, $4<m<6$, ... $2d\!-\!2<m<2d$ have so far been missing.

Note that by (\ref{1.2}) existence of a nontrivial lower bound on $|H(m)|$
for $|1-m|<\sqrt{1-c_1\,\e}$ is equivalent to 
the Wilson-Dirac operator $D_w$ having no real eigenvalues in the open interval
$]1-\sqrt{1-c_1\,\e}\,,\,1+\sqrt{1-c_1\,\e}\,[$. To extend this to a general 
localisation result for the real eigenvalues of $D_w$ 
existence of lower bounds on $|H(m)|$ for 
$k\!-\!1<m<k\!+\!1\,$, $\ k\!=\!1,3,\dots,2d\!-\!1$ needs to be established.

Our aim in this paper is to generalise (\ref{2.1}) to bounds of the following form:
\be
|H(m)|\;\ge\;\sqrt{1-c_k\,\e}-|k-m|\quad,\qquad{}k=1,3,5,\dots,2d-1
\label{2.2}
\ee
For given $m\in\;\rb{}k\!-\!1\,,\,k\!+\!1\,\lb$ this lower bound is nontrivial 
when $\e$ in the smoothness condition (\ref{1.1}) is chosen such that 
$\e<(1-(k-m)^2)/c_k$. On the other hand, if we only require  
$\e<1/c_k$ for all $k$ then the bound is nontrivial for 
all values of $m$ except those lying in one of the following intervals:
\be
&&\lb0\;,\,1-\sqrt{1-c_1\,\e}\ \rb\,, \nonumber \\
&&\lb{}k+\sqrt{1-c_k\,\e}\ ,\;k+2-\sqrt{1-c_{k+2}\,\e}\ \rb\,,
\qquad\ \ k\!=\!1,3,\dots,2d\!-\!1\,, \nonumber \\
&&\lb2d-1+\sqrt{1-c_{2d-1}\,\e}\ ,\;2d\ \rb
\label{2.3}
\ee
illustrated in Fig. \ref{bwd1} below. In this case the real eigenvalues of $D_w$ 
(in units of $1/a$) must lie in these intervals.
Clearly when $\e$ is small these intervals are localised around the values 
$0,2,4,\dots,2d$. More specifically, we see that the real eigenvalues
of $D_w$ are guaranteed to lie in the intervals $[0,\delta]\,$, 
$[2p-\delta,2p+\delta]\,$ ($p\!=\!1,2,\dots,d-1$), $[2d-\delta,2d]$
when $\e<(1-(1-\delta)^2)/c_k$ for all $k\!=\!1,3,\dots,2d-1$. This is the
advertised localisation result on the real spectrum of $D_w$. 
Explicit values for the $c_k$'s will be determined in the next section.
\begin{figure}
$$
\epsfxsize=15cm \epsfbox{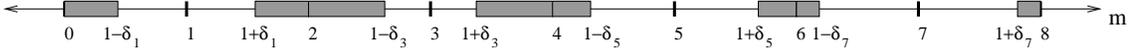}
$$
\caption{Illustration of the intervals (\ref{2.3}) (with 
$\delta_k\equiv\sqrt{1-c_k\,\e}$) in the dimension $d\!=\!4$ case. The bounds
(\ref{2.2}) imply that the real eigenvalues of $D_w$ lie in these intervals.}
\label{bwd1}
\end{figure}

\section{Derivation of the bounds}

The Wilson-Dirac operator $D_w$ with general Wilson parameter $r$ is given by
\be
aD_w(r)=\sum_{\mu}\ \gamma^{\mu}\frac{1}{2}(T_{+\mu}-T_{-\mu})
+r(1-\frac{1}{2}(T_{+\mu}+T_{-\mu}))
\label{3.1}
\ee
where $T_{\pm\mu}$ are the forward/backward parallel transporters
($(T_{+\mu})_{xy}=U_{\mu}(x)\delta_{x,y-\hat{\mu}}\,$, 
$(T_{\pm\mu})^*=(T_{\pm\mu})^{-1}=T_{\mp\mu}$). $D_w$ is an operator on the
lattice spinor fields living on a hypercubic lattice on 
an even $d$-dimensional 
Euclidean spacetime and taking values in some unitary representation of the
(unspecified) gauge group. The spacetime may be either the infinite volume 
${\bf R}^d$ or finite volume $d$-torus $T^d$. In the former case the (completion of
the) space of spinor fields is an infinite-dimensional Hilbert space, while in
the latter case it is simply a finite-dimensional complex vector space with 
inner product. In the following $||\cdot||$ denotes operator norm. 
Clearly $||T_{\pm\mu}||=1$, so $D_w$ is bounded. A well-known, important consequence 
of (\ref{1.1}) is 
\be
||\;\lb{}T_{\pm\mu}\,,\,T_{\pm\nu}\,\rb\;||\,\le\,\e\quad,
\qquad\ ||\;\lb{}T_{\pm\mu}\,,\,T_{\mp\nu}\,\rb\;||\,\le\,\e
\label{3.2}
\ee

It is useful to define the hermitian operators 
\be
S_{\mu}=\frac{1}{2i}(T_{+\mu}-T_{-\mu})\ ,\quad
\ C_{\mu}=\frac{1}{2}(T_{+\mu}+T_{-\mu})\ ,\quad\ R_{\mu}=1-C_{\mu}
\label{3.3}
\ee
These have bounds $-1\le{}S_{\mu}\le1\,$, $-1\le{}C_{\mu}\le1\,$, 
$0\le{}R_{\mu}\le2$ and satisfy (in any gauge background) the following identities:
\be
\lb{}S_{\mu}\,,\,C_{\mu}\,\rb=0\quad&,&\quad\lb{}S_{\mu}\,,\,R_{\mu}\,\rb=0
\label{3.4} \\
S_{\mu}^2+C_{\mu}^2=1\quad&,&\quad{}S_{\mu}^2=R_{\mu}(2-R_{\mu}) \label{3.5}
\ee
The Wilson-Dirac operator can then be written as 
\be
D_w(r)=\frac{1}{a}\sum_{\mu}\ i\gamma^{\mu}S_{\mu}+rR_{\mu}
\label{3.7}
\ee
For later use we also note the relations
\be
R_{\mu}=\frac{1}{2}\nabla_{\mu}^*\nabla_{\mu}\quad\qquad
2-R_{\mu}=\frac{1}{2}(2+\nabla_{\mu})^*(2+\nabla_{\mu})
\label{3.6}
\ee
where $\nabla_{\mu}=T_{+\mu}-1$.

The Hermitian Wilson-Dirac operator (normalised by $1/a$) is given by 
\be 
H(m,r)=\g5(aD_w(r)-rm)
\label{3.8}
\ee
$|H(m,r)|=\sqrt{H(m,r)^2}$ is defined via spectral theory.
In the following we set $r=1$ and consider $H(m)=H(m,1)$; the case of general
$r$ is dealt with in the Appendix.

To derive the desired bounds (\ref{2.2}) it suffices to show the following:
\be
H(k)^2\;\ge\;1-c_k\e\quad,\quad\ k\!=\!1,3,\dots,2d\!-\!1
\label{3.9}
\ee
Indeed, the eigenvalues $\lambda(m)$ of $H(m)$ satisfy $|\frac{d\lambda}{dm}|\le1$
\cite{Kerler,Neu(bound)}, implying $|H(m')|\ge|H(m)|-|m-m'|$ (an alternative derivation 
of this had also been given in the first paper of \cite{L(local)}),
and this together with (\ref{3.9}) implies the bounds (\ref{2.2}).

To derive bounds of the form (\ref{3.9}) we use (\ref{3.5})--(\ref{3.6}) to
express $H(m)^2$ as follows:
\be
H(m)^2&=&(aD_w^*-m)(aD_w-m) \nonumber \\
&=&1+\chi(m)+E'
\label{3.10}
\ee
where
\be
\chi(m)&=&\sum_{\mu}S_{\mu}^2+(-m+\sum_{\mu}R_{\mu})^2-1 \label{3.12} \\
&=&\sum_{\mu\ne\nu}R_{\mu}R_{\nu}-2(m-1)\sum_{\mu}R_{\mu}+m^2-1 \label{3.13}
\ee
and 
\be
E'=\sum_{\mu\ne\nu} \gamma^{\mu}\gamma^{\nu}\frac{1}{2}\lb{}S_{\mu}\,,\,S_{\nu}\,\rb
+i\gamma^{\mu}\lb{}S_{\mu}\,,\,C_{\nu}\,\rb
\label{3.11}
\ee
Using (\ref{3.2}) and triangle inequalities a bound on $E'$ of the form
\be
||E'||\;\le\;c'\e
\label{3.14}
\ee
can be obtained. The value for $c'$ obtained in \cite{Neu(bound)} in the 4-dimensional
case is $c'=6(1+\sqrt{2})\approx14.5$ and generalises to $c'=(1+\sqrt{2})d(d-1)/2$
in $d$ dimensions.

To complete the derivation of (\ref{3.9}) we need to show that $\chi(k)$ can be
written in the form 
\be
&&\chi(k)=P(k)+E(k)\ ,\quad\ P(k)\,\ge\,0\ ,\quad\ ||E(k)||\,\le\,c_k''\e\quad
\ \mbox{for}\ \ k\!=\!1,3,\dots,2d\!-\!1 \nonumber \\
&&\label{3.15}
\ee
It then follows from (\ref{3.10}) that (\ref{3.9}) is satisfied with 
$c_k=c_k''+c'$.

It is easy to derive a decomposition and bound (\ref{3.15}) in the $k=1$ case
\cite{L(local),Neu(bound)}. In this case (\ref{3.13}) reduces to
\be
\chi(1)=\sum_{\mu\ne\nu}R_{\mu}R_{\nu}\,.
\label{3.16}
\ee
Using (\ref{3.6}) one finds $R_{\mu}R_{\nu}=
\frac{1}{2}\nabla_{\mu}^*\nabla_{\mu}R_{\nu}=
\frac{1}{2}\nabla_{\mu}^*R_{\nu}\nabla_{\mu}+\frac{1}{2}\nabla_{\mu}^*
[\nabla_{\mu}\,,\,R_{\nu}\,]=P_{\mu\nu}+E_{\mu\nu}$ where 
$P_{\mu\nu}=\frac{1}{2}\nabla_{\mu}^*R_{\nu}\nabla_{\mu}\ge0$ and 
$||E_{\mu\nu}||\le\e$, leading to $\chi(1)=P(1)+E(1)$ with $P(1)\ge0$ and
$||E(1)||\le{}d(d-1)\e$ in $d$ dimensions \cite{L(local)}.
A more subtle decomposition $\chi(1)=P(1)+E(1)$ was derived in \cite{Neu(bound)}
for which $||E(1)||\le\frac{1}{2}d(d-1)\e$. In this way the $k=1$ bound 
(\ref{2.1}) was obtained with $c_1=c_1''+c'=6+6(1+\sqrt{2})\approx20.5$ in
4 dimensions \cite{Neu(bound)}.

Our goal now is to derive a decomposition and bound (\ref{3.15}) for $\chi(k)$
in the case of general $k=1,3,\dots,2d-1$. Setting 
\be
R_{\mu}^{(0)}=2-R_{\mu}\qquad\mbox{and}\qquad{}R_{\mu}^{(1)}=R_{\mu}
\label{3.17}
\ee
we begin by noting the identity
\be
\chi(m)=\tchi(m)+\tchi(m)_{rev}
\label{3.18}
\ee
where 
\be
\tchi(m)=\frac{1}{2^{d+1}}\sum_{q_1,\dots,q_d=0,1}((m-2(q_1+\dots+q_d))^2-1)
R_1^{(q_1)}R_2^{(q_2)}\cdots{}R_d^{(q_d)}
\label{3.19}
\ee
$\tchi(m)_{rev}$ is defined by replacing $R_1^{(q_1)}R_2^{(q_2)}\cdots{}R_d^{(q_d)}$
by $R_d^{(q_d)}\cdots{}R_2^{(q_2)}R_1^{(q_1)}$ in (\ref{3.19}).
The key feature of this expression is that, unlike the original expression
(\ref{3.13}), it is a sum of monomials in the positive operators $R_{\mu}$ 
and $2-R_{\nu}$ (recall $0\le{}R_{\mu}\le2$) with {\em positive coefficients}
when $m$ is an odd integer (in particular when $m\!=\!k\!=\!1,3,\dots,2d\!-\!1$).
As we will see shortly, this provides for a decomposition $\chi(k)=P(k)+E(k)$
of the form required in (\ref{3.15}).

To derive (\ref{3.18}), consider the expansion of $\tchi(m)$ in powers of
the $R_{\mu}$'s:
\be
\tchi=\alpha_0+\alpha_1\sum_{\mu}R_{\mu}+\dots+\alpha_p\sum_{\mu_1<\cdots<\mu_p}
R_{\mu_1}\cdots{}R_{\mu_p}+\dots+\alpha_dR_1R_2\cdots{}R_d
\label{3.20}
\ee
The expansion of $\tchi(m)_{rev}$ is identical except that the ordering of the
$R_{\mu}$'s is reversed. In light of (\ref{3.13}), to derive (\ref{3.18}) it
suffices to show that 
\be
\alpha_0=\frac{1}{2}(m^2-1)\ ,\quad\alpha_1=-(m-1)\ ,\quad\alpha_2=1
\quad\mbox{and}\ \ \alpha_p=0\ \ \mbox{for}\ \ p\ge3
\label{3.21}
\ee
Let us focus on the term of order $p$ in (\ref{3.20}). It gets contributions from
the terms in (\ref{3.19}) with $q_1+\dots+q_d\le{}p$. The terms with 
$q_1+\dots+q_d=s$ are 
\be
\frac{1}{2^{d+1}}((m-2s)^2-1)\sum_{\nu_1<\cdots<\nu_s}
(2-R_1)\cdots(2-R_{\nu_1-1})R_{\nu_1}(2-R_{\nu_1+1})\cdots&&\nonumber \\
\qquad\qquad\qquad(2-R_{\nu_s-1})R_{\nu_s}(2-R_{\nu_s+1})\cdots(2-R_d)&& 
\label{3.22}
\ee
For $s\le{}p$ the contribution of this to the $\alpha_p$-term in (\ref{3.20}) is
\be
\frac{1}{2^{d+1}}((m-2s)^2-1)\,(-1)^{p-s}\,2^{d-p}\left\lb
{\textstyle {p \atop s}}\right\rb
\sum_{\mu_1<\cdots<\mu_p}R_{\mu_1}\cdots{}R_{\mu_p}
\label{3.23}
\ee
(the binomial coefficient $\left\lb{p \atop s}\right\rb$ appears because it is the
number of ways to pick $s$ distinct elements from a set of $p$ elements).
It follows that
\be
\alpha_p=\frac{1}{2^{p+1}}\sum_{s=0}^p((m-2s)^2-1)\left\lb
{\textstyle {p \atop s}}\right\rb
\,(-1)^{p-s}
\label{3.24}
\ee
From this we find $\alpha_0=\frac{1}{2}(m^2-1)\,$, $\alpha_1=-(m-1)$ and
$\alpha_2=1$ as claimed in (\ref{3.21}). In the $p\ge3$ case we calculate
\be
\alpha_p&=&\frac{1}{2^{p+1}}\sum_{s=0}^p(4s^2-4ms+m^2-1)\left\lb
{\textstyle {p \atop s}}
\right\rb\,(-1)^{p-s} \nonumber \\
&=&\frac{1}{2^{p+1}}\bigg(\,4p(p-1)\sum_{s=0}^{p-2}\left\lb
{\textstyle {p-2 \atop s}}\right\rb\,(-1)^{p-2-s}\,+\,
4(m-1)p\sum_{s=0}^{p-1}\left\lb{\textstyle {p-1 \atop s}}\right\rb\,(-1)^{p-1-s}
\nonumber \\
&&\qquad\qquad+\,(m^2-1)\sum_{s=0}^p\left\lb
{\textstyle {p \atop s}}\right\rb\,(-1)^{p-s}\,\bigg) \nonumber \\
&=&0
\label{3.25}
\ee
(each sum vanishes since $\sum_{s=0}^{p-2}\left\lb{p-2 \atop s}\right\rb\,
(-1)^{p-2-s}=(1-1)^{p-2}$ etc.). This completes the derivation of (\ref{3.21}),
thereby establishing (\ref{3.18}).

We now show how (\ref{3.18})--(\ref{3.19}) leads to a decomposition 
$\chi(k)=P(k)+E(k)$ of the form (\ref{3.15}).
The operator product $R_1^{(q_1)}\cdots{}R_d^{(q_d)}$ in (\ref{3.19}) decomposes
into a positive piece and a piece involving commutators as follows. Set
\be
\nabla_{\mu}^{(0)}=2+\nabla_{\mu}=T_{+\mu}+1\quad\mbox{and}\quad
\nabla_{\mu}^{(1)}=\nabla_{\mu}=T_{+\mu}-1
\label{3.26}
\ee
then $||\nabla_{\mu}^{(q)}||\le2$ for $q=0,1$ and, by (\ref{3.6}) and (\ref{3.17}),
\be
R_{\mu}^{(q)}=\frac{1}{2}(\nabla_{\mu}^{(q)})^*\nabla_{\mu}^{(q)}\,.
\label{3.27}
\ee
Using this and commutator relations 
$[O,O_1\cdots{}O_p]=\sum_{s=1}^pO_1\cdots{}O_{s-1}[O,O_s]O_{s+1}\cdots{}O_p$
we obtain
\be
R_1^{(q_1)}\cdots{}R_d^{(q_d)}=P^{(q_1,\dots,q_d)}+E^{(q_1,\dots,q_d)}
\label{3.28}
\ee
with
\be
P^{(q_1,\dots,q_d)}&=&\frac{1}{2^d}(\nabla_d^{(q_d)}\cdots\nabla_1^{(q_1)})^*
\,\nabla_d^{(q_d)}\cdots\nabla_1^{(q_1)} \label{3.29} \\
E^{(q_1,\dots,q_d)}&=&\sum_{p=1}^{d-1}\frac{1}{2^p}
(\nabla_p^{(q_p)}\cdots\nabla_1^{(q_1)})^* \nonumber \\
&&\quad\times\,\bigg(\,\sum_{s=p}^dR_p^{(q_p)}\cdots{}R_{s-1}^{(q_{s-1})}
\,\lb\nabla_p^{(q_p)}\,,\,R_s^{(q_s)}\,\rb\,R_{s+1}^{(q_{s+1})}\cdots{}R_d^{(q_d)}
\bigg)\nabla_{p-1}^{(q_{p-1})}\cdots\nabla_1^{(q_1)} \nonumber \\
&&\label{3.30}
\ee
Clearly $P^{(q_1,\dots,q_d)}\ge0$. Furthermore, the bounds 
$||\nabla_{\mu}^{(q_{\mu})}||\,,\;||R_{\mu}^{(q_{\mu})}||\le2$ and, by
(\ref{3.2}) $||\,[\nabla_{\mu}^{(q_{\mu})},\,R_{\nu}^{(q_{\nu})}\,]\,||\le\e$,
together with triangle inequalities, lead to the bound
\be
||E^{(q_1,\dots,q_d)}||\;\le\;c\,\e
\label{3.31}
\ee
where
\be
c=\sum_{p=1}^{d-1}\sum_{s=p}^d2^{d-p-1}\,2^{p-1}=2^{d-3}\,(d-1)\,(d+2)
\label{3.32}
\ee
The reversed product $R_d^{(q_d)}\cdots{}R_1^{(q_1)}$ has an analogous
decomposition $P_{rev}^{(q_1,\dots,q_d)}+E_{rev}^{(q_1,\dots,q_d)}$ with identical
bounds. Consequently, by (\ref{3.18})--(\ref{3.19}) we get a decomposition
\be
\chi(m)=P(m)+E(m)
\label{3.34}
\ee
where $P(m)$ and $E(m)$ are given by (\ref{3.19}) with 
$R_1^{(q_1)}\cdots{}R_d^{(q_d)}$ replaced by 
$P^{(q_1,\dots,q_d)}+P_{rev}^{(q_1,\dots,q_d)}$ and 
$E^{(q_1,\dots,q_d)}+E_{rev}^{(q_1,\dots,q_d)}$, respectively. The coefficient in 
the summand in (\ref{3.19}) is $\ge0$ when $m$ is an odd integer, hence
$P(k)\ge0$ for odd $k$ and in particular for $k=1,3,\dots,2d-1$ as required in
(\ref{3.15}). Furthermore, from (\ref{3.31})--(\ref{3.32}) we get a bound
\be
||E(k)||\;\le\;c_k''\,\e
\label{3.35}
\ee
with
\be 
c_k''&=&2c\bigg(\,\frac{1}{2^d}\sum_{q_1,\dots,q_d=0,1}
((k-2(q_1+\dots+q_d))^2-1)\bigg) \nonumber \\
&=&\frac{2c}{2^d}\sum_{p=0}^d\left\lb{\textstyle {d \atop p}}\right\rb\,
((k-2p)^2-1) \nonumber \\
&=&2^{d-3}\,(d-1)\,(d+2)\,((k-d)^2-1+d)\qquad\quad\,(d\,\ge\,2)
\label{3.36}
\ee
Thus we have established the existence of a decomposition and bound (\ref{3.15})
for $\chi(k)$ for general $k=1,3,\dots,2d-1$. By our previous discussion this
implies the existence of the desired bounds (\ref{2.2}). We remark that 
(\ref{3.36}) is invariant under $k\to2d-k$. This is as expected in light of the
well-known fact that a lower bound on $|H(m)|$ is also a lower bound on
$|H(2d-m)|$ (see the Appendix).

The bound (\ref{3.35})--(\ref{3.36}) is rather weak. For example, in the $d=4$
case it is
\be
c_k''(d=4)\,=\,36((k-4)^2-1)+144\,,
\label{3.37}
\ee
giving in the $k=1$ case $c_1''=432$, which is much larger than the values
$c_1''=12$ and $c_1''=6$ obtained in \cite{L(local)} and \cite{Neu(bound)},
respectively. Note however that for the applications discussed in this paper
it suffices simply to show the existence of bounds of the form (\ref{2.2})
without necessarily finding sharp ones. The largeness of $c_k''$ in the above
bound is due to the large number of terms in the expression (\ref{3.19}) for
$\tchi(m)$. In practice it is often possible to simplify this expression
such that a sharper bound (i.e. smaller $c_k''$) can be derived. We discuss this
in the $d=4$ case in the following.

In the remainder of this section we specialise to dimension $d=4$ and consider
$\chi(k)$ for $k=1,3,5,7$. We wish to simplify the expression 
(\ref{3.18})--(\ref{3.19}) for $\chi(k)$ in order to get bounds with smaller 
$c_k''$. In order to have a decomposition $\chi(k)=P(k)+E(k)$ the simplified
expression must continue to be a sum of monomials in the positive operators
$R_{\mu}\,$, $(2-R_{\nu})$ with positive coefficients. In the $k\!=\!1$ case 
(\ref{3.18}) simplifies to $\chi(1)=\sum_{\mu\ne\nu}R_{\mu}R_{\nu}$ (recall
(\ref{3.16})) from which the previously discussed bounds with 
$c_1''=12$ \cite{L(local)} and $c_1''=6$ \cite{Neu(bound)} can be derived.
In the $k\!=\!7$ case $\tchi(7)$ reduces to $\sum_{\mu<\nu}(2-R_{\mu})(2-R_{\nu})$,
leading to 
\be
\chi(7)=\sum_{\mu\ne\nu}(2-R_{\mu})(2-R_{\nu})
\label{3.38}
\ee
Arguments analogous to the ones in \cite{L(local)}, and \cite{Neu(bound)},
lead to bounds with $c_7''=c_1''=12$, and $c_7''=c_1''=6$, respectively.
Turning now to the $k\!=\!3$ case, (\ref{3.19}) gives
\be
\tchi(3)&=&{\textstyle \frac{1}{4}}(2-R_1)(2-R_2)(2-R_3)(2-R_4) 
+{\textstyle \frac{3}{4}}R_1R_2R_3R_4 \nonumber \\
&&+\,{\textstyle \frac{1}{4}}\Big((2-R_1)R_2R_3R_4+R_1(2-R_2)R_3R_4 
+R_1R_2(2-R_3)R_4 \nonumber \\
&&\qquad\quad+R_1R_2R_3(2-R_4)\Big) 
\label{3.39}
\ee
In this case there does not appear to be a major simplification with the required 
properties. In fact it is quite easy to show that $\chi(3)$ cannot be written as
a sum of monomials of order $\le3$ in $R_{\mu}\,$, $(2-R_{\nu})$ with positive
coefficients (we leave this as an exercise for the reader). Minor simplifications
are possible though, for example:
\be
\tchi(3)&=&{\textstyle \frac{1}{4}}((2-R_1)(2-R_2)(2-R_3)(2-R_4)
+R_1R_2R_3(2-R_4)) \nonumber \\
&&+\,{\textstyle \frac{1}{2}}(R_1R_2R_4+R_1R_3R_4+R_2R_3R_4)
\label{3.40}
\ee
$\tchi(3)_{rev}$ simplifies analogously. Estimates of the kind used to derive 
(\ref{3.31})--(\ref{3.32}) show that the decomposition $P(3)+E(3)$ of the
resulting expression for $\chi(3)$ satisfies $||E(3)||\le{}c_3''\,\e$ with
$c_3''=42$. This is considerably smaller than the
value $c_3''=144$ provided by (\ref{3.37}). It is plausible that a bound with
even smaller $c_3''$ can be derived, e.g. by an extension of the arguments of
\cite{Neu(bound)}, but we will not pursue this here. 
Finally, in the $k=5$ case analogous arguments lead (as expected) to a bound with
$c_5''=c_3''=42$ (we omit the details).

\section{Heuristic considerations}

In this section we present a heuristic argument which provides a clearer intuitive
understanding of why bounds of the form derived in the previous section should 
hold. We go on to heuristically derive certain properties of the spectral flow 
of $H(m)$ previously observed in numerical studies (e.g. \cite{Itoh,Edwards}).

Consider a ``near zero-mode'' for $H(m)$:
\be
H(m)^2\psi\,\approx\,0
\label{4.1}
\ee
If $\e$ in the smoothness condition (\ref{1.1}) is small then $E'\approx0$ in
(\ref{3.10}), and (\ref{4.1}) becomes (recall $C_{\mu}=1-R_{\mu}$)
\be
\bigg(\,\sum_{\mu}S_{\mu}^2+(-m+\sum_{\mu}(1-C_{\mu}))^2\,\bigg)\,\psi\,
\approx\,0\,.
\label{4.2}
\ee
Since $S_{\mu}^2\ge0$ it follows that $S_{\mu}\psi\approx0$ for $\mu=1,\dots,2d$
and consequently, by (\ref{3.5}), $C_{\mu}^2\,\psi=(1-S_{\mu}^2)\psi\approx\psi$
which implies that $C_{\mu}\psi\approx(-1)^{j_{\mu}}\psi$ for 
$j_{\mu}\!=\!0$ or $1$. Then (\ref{4.2}) reduces to
\be
0&\approx&\Big(\,-m+\sum_{\mu}(1-C_{\mu})\Big)^2\,\psi \nonumber \\
&\approx&\Big(\,-m+\sum_{\mu}(1-(-1)^{j_{\mu}})\Big)^2\,\psi
\label{4.3}
\ee
which implies
\be
m\,\approx\,\sum_{\mu}(1-(-1)^{j_{\mu}})\,.
\label{4.4}
\ee
Thus we see heuristically that when $\e$ is small the only values of $m$ for
which $H(m)$ can have ``near zero-modes'' are $m=0,2,4,\dots,2d$.
This makes plausible the result of the previous section, namely that when $m$
is away from these values a nonzero lower bound on $|H(m)|$ should exist.

In fact the above heuristic approach can be further developed to get an alternative
rigorous derivation of the bounds (\ref{2.2}) \cite{DA(unpublished)}.
However, the argument is technically more complicated than the one in \S3, and does
not lead to sharper bounds, so we do not present it here.

We now proceed to study the spectral flow of $H(m)$. For this it is useful to
introduce the operators $\T_{\mu}$ defined by\footnote{These have proved useful in
previous lattice fermion contexts; see, e.g., \cite{Fuji} and the ref.'s therein.}
\be
(\T_{\mu})_{xy}=\g5\gamma^{\mu}(-1)^{n_{\mu}}\delta_{xy}\qquad\qquad
\ (n_{\mu}=x_{\mu}/a\in{\bf Z})
\label{4.5}
\ee
These have the following properties: $\T_{\mu}^2\!=\!-1\,$, 
$\ \T_{\mu}\T_{\nu}\!=\!-\T_{\nu}\T_{\mu}$ for $\mu\ne\nu\,$, 
$\ [\T_{\mu}\,,\,\gamma^{\nu}S_{\nu}\,]\!=\!0\,$, 
$\ C_{\mu}\T_{\mu}\!=\!-\T_{\mu}C_{\mu}\,$, $\ [\T_{\mu}\,,\,C_{\mu}\,]\!=\!0$ 
for $\mu\ne\nu$, and $\T_{\mu}\g5\!=\!-\g5\T_{\mu}$.
Using these we find
\be
H(m)\T_{\mu}=-\T_{\mu}(H(m)+2\g5C_{\mu})\qquad\qquad\ \mbox{(no sum over $\mu$)}
\label{4.6}
\ee
By (\ref{3.2}) $[H(m)\,,\,C_{\mu}\,]\approx0$ when $\e$ is small, so the eigenspaces
of $H(m)$ can be decomposed into approximate eigenspaces for the $C_{\mu}$'s.
I.e. for eigenvectors $\psi(m)$ of $H(m)$ with $H(m)\psi=\lambda(m)\psi(m)$
we can assume $C_{\mu}\psi(m)\approx{}c_{\mu}\psi(m)$. The eigenvalues $c_{\mu}$
are independent of $m$ since $C_{\mu}$ is independent of $m$ and has discrete 
spectrum. Then, by(\ref{4.6}),
\be
H(m)\T_{\mu}\,\psi(m)=-\T_{\mu}H(m-2c_{\mu})\,\psi(m)\qquad\qquad
\mbox{(no sum over $\mu$)}
\label{4.7}
\ee
Set $\psi_{\mu}(m):=\T_{\mu}\psi(m-2c_{\mu})$. It follows from (\ref{4.7}) that 
$\psi_{\mu}(m)$ is an approximate eigenvector for $H(m)$ with eigenvalue 
$\approx-\lambda(m-2c_{\mu})$. Similarly we find
\be
H(m)\,\psi_{\mu_1\cdots\mu_p}(m)\;\approx\;(-1)^p\,
\lambda(m-2(c_{\mu_1}+\dots{}c_{\mu_p}))\,\psi_{\mu_1\cdots\mu_p}(m)
\label{4.8}
\ee
where $\psi_{\mu_1\cdots\mu_p}(m):=\T_{\mu_1}\cdots\T_{\mu_p}\,
\psi(m-2(c_{\mu_1}+\dots+c_{\mu_p}))$, when the $\mu_j$'s are all mutually distinct. 

\begin{figure}
$$
\epsfxsize=15cm \epsfbox{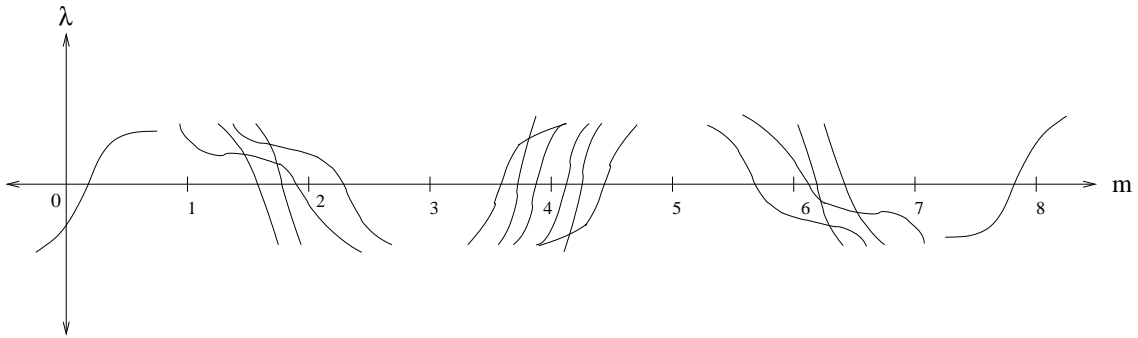}
$$
\caption{Illustration of the spectral flow associated with a typical eigenvector family
of $H(m)$ of the kind discussed in the text in dimension $d\!=\!4$: A crossing near
$m\!=\!0$ with $+$sign is associated with 4 crossings near $m\!=\!2$ with $-$sign, 
6 crossings near $m\!=\!4$ with $+$sign, 4 crossings near $m\!=\!6$ with
$-$sign and 1 crossing near $m\!=\!8$ with $+$sign.}
\label{bwd2}
\end{figure}

Now, if $\lambda(m)$ crosses zero near $m=0$ then by our previous argument (recall 
(\ref{4.4}))
$c_{\mu}\approx(-1)^{j_{\mu}}$ with $\sum_{\mu}(1-(-1)^{j_{\mu}})\approx0$, i.e. 
$c_{\mu}\approx1$ for all $\mu\,$, and (\ref{4.8}) becomes
\be
H(m)\,\psi_{\mu_1\cdots\mu_p}(m)\;\approx\;(-1)^p\,\lambda(m-2p)\,
\psi_{\mu_1\cdots\mu_p}(m)
\label{4.9}
\ee
i.e. $\psi_{\mu_1\cdots\mu_p}(m)$ is an approximate eigenvector for $H(m)$ whose 
approximate eigenvalue $\lambda(-1)^p\lambda(m-2p)$ crosses zero near $m=2p$. 
Furthermore, the sign of the crossing is $(-1)^p$ relative to the sign of the crossing 
of zero by $\lambda(m)$ near $m=0$.
We note the following: (i) If $\{\mu_1,\dots,\mu_p\}\ne\{\nu_1,\dots,\nu_p\}$ then 
$\psi_{\mu_1\cdots\mu_p}(m)$ and $\psi_{\nu_1\cdots\nu_p}(m)$ are approximately 
orthogonal since they are approximate eigenvectors for the $C_{\mu}$'s with different 
eigenvalues.
(ii) $\psi_{\mu_1\cdots\mu_p}(m)$ is unchanged up to a sign under a change of ordering 
of the $\mu_j$'s (since $\T_{\mu}\T_{\nu}=-\T_{\nu}\T_{\mu}$ for $\mu\ne\nu$). 
Hence we can assume that the $\mu_j$'s are ordered so that $\mu_1<\dots<\mu_p$.
(iii) If $\tpsi(m)$ is an eigenvector for $H(m)$ whose eigenvalue $\tlambda(m)$ crosses 
zero at some value $m_0$ then by (\ref{4.4}) $m_0\approx\sum_{\mu}(1-(-1)^{j_{\mu}})$ 
where $C_{\mu}\,\tpsi(m)\approx(-1)^{j_{\mu}}\,\tpsi(m)$ for $j_{\mu}\!=\!0$ or $1$. 
Any such eigenvector arises in the way described above, i.e. 
$\tpsi(m)=\psi_{\mu_1\cdots\mu_p}(m)=\T_{\mu_1}\cdots\T_{\mu_p}\,\psi(m-2p)$.
Indeed, we set $\psi(m)=(-1)^p\T_{\mu_1}\cdots\T_{\mu_p}\tpsi(m+2p)$ with the $\mu_j$'s 
being the $\mu$'s for which $j_{\mu}\!=\!1$. Then by (\ref{4.8}) $\psi(m)$ is an 
approximate eigenvector for $H(m)$ whose eigenvalue $\lambda(m)$ is $\approx0$ at 
some value of $m$ near zero. (To see this, recall $\T_{\mu}^2=-1$.) Thus we have 
heuristically established the following:
{\em The eigenvectors of $H(m)$ whose eigenvalues cross zero at some value of $m$ 
can be naturally grouped into sets of $2^d$ elements: One of the eigenvectors, 
$\psi(m)$, has eigenvalue $\lambda(m)$ crossing zero near $m=0$ with crossing sign 
$\pm$. There are $d$ eigenvectors, $\psi_d(m)$, with eigenvalues crossing zero near 
$m=2$ with sign $\mp$, and more generally $d!/(p!(d-p)!)$ eigenvectors 
$\psi_{\mu_1\cdots\mu_p}(m)\,$, $\mu_1<\dots<\mu_p\,$, with eigenvalues 
crossing zero near $m=2p$ with crossing sign $\mp(-1)^p$ for $p=1,2,\dots,d$.} 
This is precisely the spectral flow property of $H(m)$ found in numerical studies 
in 2 and 4 dimensions \cite{Itoh,Edwards}.
An illustration of the spectral flow associated with one such family in the
$d\!=\!4$ case is given in Fig. \ref{bwd2}. 
The Hermitian Wilson-Dirac operator in any gauge background $U$ has the well-known 
property $H(U,m)=-H(-U,2d-m)$, so that $H(m)=OH(U,2d-m)O^{-1}$ for a certain unitary
operator $O$ (see, e.g., \cite{Gatt,Edwards}).\footnote{For $O$ to exist
the number of lattice sites along each edge of $T^d$ must be even.} 
Hence if $\lambda(m)$ is an eigenvalue for $H(m)$ then $-\lambda(m)$ is an eigenvalue
for $H(2d-m)$. This property must be manifested in the eigenvalues of the family of 
eigenvectors of $H(m)$ discussed above, and is also illustrated in Fig. \ref{bwd2}.
Combining this spectral property of $H(m)$ with the fact that the index of the
overlap Dirac operator equals $-1/2$ times the spectral asymmetry of $H(m)$ 
\cite{ov,Neu}, an immediate consequence is the relation
$index(D_{ov}(m))=-index(D_{ov}(2d-m))$ which was emphasised in \cite{Chiu}.

\section{Applications of the bounds}

We have already seen in \S2 how the general bounds (\ref{2.2}) lead to a localisation 
result on the real spectrum of the Wilson-Dirac operator, thus providing an analytic
understanding of the numerical results on the real spectrum in ``smooth'' gauge 
backgrounds. In this section we discuss applications of the bounds to overlap
fermions \cite{Neu}.
The general bounds allow analytic results on the overlap Dirac operator $D_{ov}$ which 
were previously derived for the $0<m<2$ to be extended to the general $m$ case 
($m\ne0,2,4,\dots,2d$). Although $0<m<2$ is the physically relevant case (i.e the
case where $D_{ov}$ is free from spurious fermion species) this restriction appears quite 
unnatural and it is of some theoretical interest to know the properties of $D_{ov}$
in the regions $2p<m<2p+2\,$, $\,p\!=\!1,2,\dots,d$ where the extra fermion species are 
present.

\vspace{1ex}

\noindent (i) {\em Locality and smooth gauge 
field-dependence of the overlap Dirac operator.}
 
With the bounds (\ref{2.2}) 
the arguments of ref.\cite{L(local)} for the locality of $D_{ov}$ and its smooth 
dependence on the lattice gauge field carry over unchanged from the $0<m<2$ case to the 
$k-1<m<k+1$ case ($k\!=\!1,3,\dots,2d-1$) after choosing $\e<(1-(k-m)^2)/c_k$ so that
the lower bound on $|H(m)|$ is greater than zero. The size of the exponential decay 
constant in the locality bound for $D_{ov}$ depends on the size of $c_k\,$, but for the 
existence of the locality bound it is enough to know that (\ref{2.2}) holds for a 
specific value of $c_k$ which is independent of the lattice gauge field. 

\vspace{1ex}

\noindent (ii) {\em Evaluation of the classical continuum limit of the axial anomaly and 
index of the overlap Dirac operator.}

The rigorous evaluation of the classical continuum limit of the axial 
anomaly\footnote{It may sound contradictory to speak of the ``classical'' continuum
limit of a purely quantum quantity such as the axial anomaly, so let us explain the
meaning: ``Classical'' refers to the fact that one considers the $a\to0$ limit of
the axial anomaly with the lattice gauge field given by the lattice transcript of a
{\em smooth continuum gauge field}. See \cite{DA(AP),DA(JMP)} for the details.} and 
index of the overlap Dirac operator at general values of $m$ requires existence of a 
nontrivial lower bound on $|H(m)|$ when the lattice is sufficiently fine 
\cite{DA(AP),DA(JMP)}. We claimed in \cite{DA(AP),DA(JMP)} that such bounds exist
and promised to provide them in a forthcoming paper. The present paper delivers on 
that promise. Again, the actual values of the $c_k$'s do not matter: The
lattice transcript of a smooth continuum gauge field automatically satisfies the
smoothness condition (\ref{1.1}) for any $\e>0$ when the lattice is sufficiently fine 
(see \cite{DA(JMP)} for the rigorous justification of this point), so all
that matters for the classical continuum limit calculations is that the bounds hold
for some choice of $c_k$'s which are independent of the gauge field and lattice spacing.

\vspace{1ex}

\noindent (iii) {\em Topological phase structure of the overlap Dirac operator.}

In the finite volume $d$-torus case
the index of $D_{ov}=\frac{1}{a}(1+\g5H(m)/|H(m)|)$ is a well-defined 
integer; it is locally constant in $m$ but may jump at the values at which $H(m)$ 
has zero-mode(s).
Thus $D_{ov}$ has different topological phases and the value of $m$ should be
chosen so that $D_{ov}$ is in the ``correct'' phase. This issue has previously been 
studied both analytically and numerically in \cite{Chiu}and numerically in \cite{Sch}. 
However, the analytic arguments in \cite{Chiu}
are problematic since they involve treating topologically nontrivial 
fields as perturbations of the trivial gauge field $U\!=\!1$.
On the other hand, the bounds (\ref{2.2}) provide rigorous nonperturbative 
insight into the topological phase structure when the lattice gauge fields are required
to satisfy the smoothness condition (\ref{1.1}) with $\e<1/c_k$ for all 
$k=1,3,\dots,2d-1\,$: they imply that there are distinct 
topological phases for $D_{ov}\,$, with each phase characterised by
$m$ being in one of the open intervals $\rb{}k-\sqrt{1-c_k\,\e}\,,\,
k+\sqrt{1-c_k\,\e}\,\lb$. The result of \cite{DA(JMP)} states that for SU($N$) gauge
fields on the $d$-torus ($d\!=\!2n\,$, $n>1$), or U(1) gauge fields on the 2-torus,
$index(D_{ov})$ coincides with the index of the continuum Dirac operator
in the classical continuum limit provided $0<m<2$.\footnote{This was shown in 
\cite{DA(JMP)} in the case of the 4-torus, but the argument generalises 
straightforwardly to the general $d\!=\!2n$-torus.} This indicates that the ``proper''
topological phase for $D_{ov}$ is the one where $m$ is in the interval 
$\rb{}1-\sqrt{1-c_1\,\e}\,,\,1+\sqrt{1-c_1\e}\,\lb$. We denote $index(D_{ov})$ by 
$Q$ when $D_{ov}$ is in this phase. A complete description of the topological phases 
for $D_{ov}$ when the smoothness condition (\ref{1.1}) is imposed is now as follows.
For $m\le0\,$, $\,D_{ov}$ is in a topologically trivial phase (i.e. $index(D_{ov})=0$
in any gauge background) \cite{ov}.
For $0<m\le1-\sqrt{1-c_1\,\e}\,$, $\,D_{ov}$ is not in a 
distinct topological phase: $index(D_{ov})$ can be any value from $0$ to $Q$
depending on the background gauge field. In a given gauge background, as $m$ is
increased from $0$ to $1-\sqrt{1-c_1\,\e}\,$, the total spectral flow of $H(m)$
is $Q$. This is due to the well-known fact that at each crossing of zero by an
eigenvalue of $H(m)$ the index of $D_{ov}$ changes by $\mp1$ depending on the sign 
of the crossing. For $1-\sqrt{1-c_1\,\e}<m<1+\sqrt{1-c_1\,\e}\,$, $\,D_{ov}$ is in 
the ``proper'' topological phase where $index(D_{ov})=Q$.
For $1+\sqrt{1-c_1\,\e}\,\le\,m\,\le\,3-\sqrt{1-c_3\,\e}\,$, $\,D_{ov}$ is no longer 
in a distinct topological phase and the spectral flow of $H(m)$ as $m$ increase 
through this region is $-dQ$. For $3-\sqrt{1-c_3\,\e}<m<3+\sqrt{1-c_3\,\e}\,$,
$\,D_{ov}$ is in another distinct topological phase with 
$index(D_{ov})=(1-d)Q$. The pattern continues as $m$ increases:
For $k-\sqrt{1-c_k\,\e}<m<k+\sqrt{1-c_k\,\e}\,$, $\,D_{ov}$ is in a distinct 
topological phase with $index(D_{ov})=
\Big(\,\sum_{p=0}^{(k-1)/2}(-1)^p\left\lb{d \atop p}\right\rb\Big)\,Q$. Then, as $m$
increases from $k+\sqrt{1-c_k\,\e}$ to $k+2-\sqrt{1-c_{k+2}\,\e}\,$, $\,D_{ov}$
is no longer in a distinct topological phase and the spectral flow of $H(m)$
through this region is $(-1)^{(k+1)/2}\left\lb{d \atop (k+1)/2}\right\rb\,Q$.
Finally, after $m$ has increased to $2d$ we have $index(D_{ov})=
\Big(\,\sum_{p=0}^d(-1)^p\left\lb{d \atop p}\right\rb\Big)\,Q=(1-1)^d\,Q=0$
and $D_{ov}$ is back in a topologically trivial phase which it remains in
for all $m\ge2d$.

The above description of the topological phase structure of $D_{ov}$ is compatible with
the results of previous numerical studies in 2 and 4 dimensions \cite{Chiu,Sch}. 
To put the above analytic 
argument on a completely rigorous footing a rigorous derivation of the heuristic result
of \S4 on the spectral flow of $H(m)$ is required. This remains as a problem for future 
work. We note however that further evidence for the validity of this description comes 
from the result 
of \cite{DA(JMP)}, which states that {\em in the classical continuum limit}
\be
index(D_{ov})&=&\Big(\,\sum_{p=0}^{(k-1)/2}(-1)^p
\left\lb{\textstyle {d \atop p}}\right\rb\Big)\,Q
\quad\ \mbox{for $\,k\!-\!1<m<k\!+\!1\ $ ($k\!=\!1,3,\dots,2d\!-\!1$)}  \nonumber \\
&=&0\qquad\quad\mbox{for $m\le0$ and $m\ge2d$}
\nonumber 
\ee
where $Q$ is the index of the continuum Dirac operator.

A generalisation of the overlap Dirac operator has been presented in 
\cite{Fuji-ov} and it would also be of interest to establish the topological phase 
structure of this operator. For this, bounds on the generalised Hermitian Wilson-Dirac
operator for general $m$ are needed. A bound has already been derived in
\cite{Fuji-Mata} for $0<m<2$ by a generalisation of the argument of
\cite{Neu(bound)}. It is plausible that bounds for general $m$ can be derived
by a generalisation of the argument in the present paper. We leave this as a potential
topic for future work.

\vspace{1ex}

\noindent (iv) {\em Instanton size-dependence of the real spectrum of the Wilson-Dirac
operator in an instanton background.}

Approximate instantons on the lattice can be obtained either through a cooling
procedure \cite{cooling}, or by taking an appropriate lattice
transcript of a continuum instanton field \cite{Teper,Edwards}. We will focus on the
latter case. In this case numerical studies have shown that the real eigenvalues
of $D_w$ are well localised around $0,2,\dots,2d$ (or equivalently, the crossings of
zero by eigenvalues of $H(m)$ occur close to these values) when the instanton is 
large at the scale of the lattice spacing, but become delocalised as the instanton size
is decreased \cite{Edwards}. The standard explanation of this is that instantons
which are small at the scale of the lattice spacing are not slowly varying at this scale
in the region in which they are localised, so their lattice transcripts are ``rough'' 
in this region. On the other hand, large instantons {\em are} slowly varying, so their 
lattice transcripts are ``smooth''. The bounds (\ref{2.2}) can be used to give a more 
precise version of this intuitive explanation as follows. A continuum instanton field
centred at $x^{(0)}$ has the form \cite{instanton-Forkel}
\be
A_{\mu}(x)=2\eta_{\alpha}^{\mu\nu}\frac{x_{\nu}-x_{\nu}^{(0)}}{|x-x^{(0)}|^2+\rho^2}
\,t^{\alpha}
\label{5.4}
\ee
where $\eta_{\alpha}^{\mu\nu}$ is the 't Hooft symbol, $t^{\alpha}$ are generators of the
SU(2) subgroup and the parameter $\rho$ specifies the size of the instanton.
Its curvature  is
\be
F_{\mu\nu}(x)=-4\eta_{\alpha}^{\mu\nu}\frac{\rho^2}{(|x-x^{(0)}|^2+\rho^2)^2}\,t^{\alpha}
\label{5.5}
\ee
When putting the instanton on the lattice with periodic boundary conditions it is 
important to transform (\ref{5.4}) to a singular gauge before taking the lattice 
transcript (and the lattice volume must also be sufficiently large that the 
singular gauge instanton is close to vanishing at the boundary) \cite{Teper,Edwards}. 
$||F_{\mu\nu}(x)||$ is not affected by this though 
since it is gauge invariant. From (\ref{5.5}) we see that $||F_{\mu\nu}(x)||$ 
diverges at $x^{(0)}$ in the limit of small instanton size $\rho$. 
Hence for small $\rho$ the lattice transcripted field generally 
violates the smoothness condition 
(\ref{1.1}) since $||1-U(p)||=||a^2F_{\mu\nu}(x)+O(a^3)||$ becomes large for plaquettes 
$p$ close to $x^{(0)}$. (This is assuming there is no special cancellation between
$a^2F_{\mu\nu}(x)$ and the $O(a^3)$ term; generically there is no reason to expect
such a cancellation to occur, and in particular when the lattice spacing is small 
$a^2F_{\mu\nu}(x)$ will dominate the $O(a^3)$ term.)
Then the localisation result of \S2
on the real spectrum of $D_w$ breaks down.\footnote{More precisely, the assumptions
under which the localisation was derived break down. This does not necessarily imply 
that the localisation result itself must break down, although it is not surprising that
it should do so. We can turn things around and interpret the numerical results on
the delocalisation of the real spectrum in small instanton backgrounds as indicating
that, in general, a smoothness requirement of the form (\ref{1.1}) is not only
sufficient but also a necessary requirement for the real spectrum of $D_w$ to be 
localised.} 

On the other hand, from (\ref{5.5}) we get a bound
\be
||F_{\mu\nu}(x)||\;\le\;\frac{4||\eta_{\alpha}^{\mu\nu}\,t^{\alpha}||}{\rho^2}
\label{5.6}
\ee
showing that $||F_{\mu\nu}(x)||$ vanishes uniformly in the limit of large $\rho$.
Consequently, for large $\rho$ the smoothness condition (\ref{1.1}) will generically be 
satisfied on sufficiently fine lattices, thereby guaranteeing localisation of the real 
spectrum of $D_w$ according to the result of \S2.

These considerations can be extended to more general gauge fields describing a collection
of topologically charged ``lumps'' (e.g. instanton--anti instanton configurations,
multi-instantons, instanton gases). The topological charge of a lump is given by
\be
Q_{lump}=\frac{1}{32\pi^2}\int_{lump}d^4x\,\e_{\mu\nu\sigma\rho}\tr{}F_{\mu\nu}(x)
F_{\sigma\rho}(x)\;\approx\;\pm1
\label{5.7}
\ee
If the lump size is small then $||F_{\mu\nu}(x)||$ must be large
in the lump region in order that the magnitude of the integral in (\ref{5.7}) can be 
$\approx1$. The smaller the lump is, the larger $||F_{\mu\nu}(x)||$ must be in 
the lump region. This generically leads to violation of the smoothness condition 
(\ref{1.1}) as before. 
On the other hand, if the lump size is large $||F_{\mu\nu}(x)||$ is not forced to be 
large in any particular region. Generically we can expect $||F_{\mu\nu}(x)||$ to decrease
with increasing lump size, and to vanish in the large lump limit. Then, by the
same argument as before, localisation of the real spectrum of $D_w$ will generically 
hold in gauge backgrounds describing topological lumps when all the lumps are sufficiently
large and the lattice is sufficiently fine.

\section{Summary}

We have derived general lower bounds on the magnitude of the spectrum of the Hermitian 
Wilson-Dirac operator:
\be
|H(m)|\;\ge\;\sqrt{1-c_k\,\e}-|k-m|\qquad\quad\mbox{for $k=1,3,\dots2d-1$.}
\nonumber
\ee
where $\e$ is the constraining parameter in the smoothness condition (\ref{1.1})
(and the Wilson parameter is $r\!=\!1$; the generalisation to arbitrary $r>0$ is 
given in the appendix.) Thus we have supplemented the previous bounds for the 
``physical'' case $k\!=\!1$ \cite{L(local),Neu(bound)} with bounds for the  
``doubler'' cases $k=3,5,\dots,2d-1$.  
The bounds were shown to hold with
\be
c_k&=&c'+c_k'' \nonumber \\
c'=(1+\sqrt{2})d(d-1)/2\ ,&&c_k''=2^{d-3}(d-1)(d+2)((k-d)^2-1+d) 
\nonumber
\ee
The bounds are rather weak due to the large size of $c_k''\,$,
which is due to the large number of terms in the expression (\ref{3.19}) for 
$\tchi(k)$. In practice it is often possible to get sharper bounds
(i.e. smaller $c_k''$) by considering simplified expressions for $\tchi(k)$. 
For example, in dimension $d\!=\!4$ we saw how such simplifications lead to 
bounds with $c_1''=c_7''=12$ and $c_3''=c_5''=42$.
In the $k\!=\!1$ case this is the same as the value obtained in \cite{L(local)}. 
It is plausible that bounds with even smaller $c_k''$ can be derived by an extension 
of the arguments of \cite{Neu(bound)} (which gave $c_1''=6$) but we did not pursue 
this. For the applications considered in this paper it suffices simply to show that 
bounds of the above form exist, without necessarily finding sharp ones.

As discussed in \S2, the lower bounds on $|H(m)|$ imply a localisation 
result on the real eigenvalues of the usual Wilson-Dirac operator: the eigenvalues 
of $D_w$ (in units of $1/a$) are localised around the values $0,2,4,\dots,2d$ when
$\e$ is sufficiently small. (A precise formulation of this statement was given 
in \S2.)

The bounds allow previous results on the overlap Dirac operator to be extended
from the $0<m<2$ case to general values of $m\,$ ($m\ne0,2,\dots,2d$).
This includes evaluation of the classical continuum limit of the axial anomaly
and index \cite{DA(AP),DA(JMP)}, and the results of \cite{L(local)} on locality of the 
overlap Dirac operator and its smooth dependence of the gauge field.
The bounds were also seen to imply the existence of topological phases for the 
overlap Dirac operator
when one restricts to the space of lattice gauge fields satisfying (\ref{1.1})
with $\e<1/c_k$ for all $k$. 
A complete description of the topological phase structure 
was obtained by combining the bounds with the heuristic result of \S4 on the spectral 
flow properties of $H(m)$. 

Finally, we pointed out how the bounds can be used to get a 
more precise understanding of why the real spectrum of the Wilson-Dirac operator
in an instanton background is generally localised around $0,2,\dots,2d$ when
the instanton size is large but becomes delocalised when the instanton is small
at the scale of the lattice spacing. 
(The argument also applies to more general gauge fields
describing a collection of ``topological lumps''.)
Our argument for delocalisation of the real spectrum in small 
instanton backgrounds involved an assumption, namely that, generically, the smoothness
condition (\ref{1.1}) is not only sufficient but also a {\em necessary} condition
for localisation of the spectrum. Numerical studies (e.g. \cite{Edwards}) seem to
indicate that this is the case, but it would be interesting if it could be
proved analytically. This is relevant for the issue of chiral
symmetry breaking in lattice gauge theory since it means
that the contribution to the density of near-zero eigenvalues of the Dirac
operator from gauge fields describing small topological lumps is reduced on the lattice.
Is this reduction an unwanted lattice artifact, or is it a genuinely physical
feature revealed by lattice regularisation (in the same way that lattice- and other
regularisations reveal the presence of anomalies which one would not have expected
from formal continuum considerations)?

\vspace{1ex}

\noindent {\bf Acknowledgements.} Part of this work was done during a visit to 
the National Center for Theoretical Sciences, Hsinchu, Taiwan in 1999. I thank the 
NCTS, and in particular Ting-Wai Chiu, for kind hospitality and support during the 
visit, and for stimulating discussions. I also thank Herbert Neuberger for exchanges
on bounds during the Chiral 99 workshop, Martin L\"uscher for supplying me with
the first paper in \cite{L(local)}, and Pierre van Baal for useful feedback on the
manuscript (including the suggestion to include the figures).
Part of the work was also done at Adelaide Univ.
where the author was supported by the Australian Research Council. The remainder
was done at Leiden Univ. where the author is supported by a Marie Curie
fellowship from the European Commission (contract HPMF-CT-2002-01716).

\appendix
\section*{Appendix}
\section{Bounds in the case of general Wilson parameter $r>0$}

Using (\ref{3.7})--(\ref{3.8}) a simple calculation gives \cite{ov}
\be
H(m,r)^2=H(0,r)^2-2mr^2\sum_{\mu}R_{\mu}+r^2m^2
\label{a1}
\ee
It follows that 
\be
|H(m,r)|\;\ge\;rm\qquad\quad\mbox{when $\ m\le0$}
\label{a2}
\ee
It is well-known that a lower bound on $|H(m)|$ is also a lower bound on
$|H(2d-m)|$, hence (\ref{a2}) implies
\be
|H(2d+m,r)|\;\ge\;rm\qquad\quad\mbox{when $\ m\ge0$}
\label{a3}
\ee
To see this explicitly, write $H(m,r)$ out according to the definitions
(\ref{3.1}) and (\ref{3.8}):
\be
H(m,r,U)=\g5\Big(\,rd-rm+\sum_{\mu}{\textstyle \frac{1}{2}}
(\gamma^{\mu}(T_{+\mu}-T_{-\mu})-r(T_{+\mu}+T_{-\mu}))\Big)
\label{a3.5}
\ee
It follows that
\be
H(2d-m,r,U)&=&-\g5\Big(\,rd-rm \nonumber \\
&&\qquad+\sum_{\mu}{\textstyle \frac{1}{2}}
(\gamma^{\mu}((-T_{+\mu})-(-T_{-\mu}))-r((-T_{+\mu})+(-T_{-\mu})))\,\Big)
\nonumber \\
&&\label{a4}
\ee
Since $(T_{+\mu})_{xy}=U_{\mu}(x)\delta_{x,y-\hat{\mu}}$ and 
$(T_{-\mu})_{xy}=U_{\mu}(x-\hat{\mu})^{-1}\delta_{x,y+\hat{\mu}}$ the replacement  
$T_{\pm\mu}\to-T_{\pm\mu}$ is equivalent to $U\to-U$. Hence (\ref{a4}) can be
written as 
\be
H(2d-m,r,U)=-H(m,r,-U)
\label{a5}
\ee
The operator $R_{\mu}(U)=1-\frac{1}{2}(T_{+\mu}+T_{-\mu})$ remains positive under
$U\to-U$, so the argument leading to (\ref{a2}) remains valid under this replacement
and we get $|H(2d+m,r,U)|=|H(-m,r,-U)|\ge{}rm$ for $m\ge0$ as claimed in (\ref{a3}).

It remains to derive the generalisation of the bounds (\ref{2.2}) in the general
$r$ case. In this case the relation (\ref{3.10})--(\ref{3.11}) becomes
\be
H(m,r)^2=\sum_{\mu}S_{\mu}^2+r^2(-m+\sum_{\mu}R_{\mu})^2+E'(r)
\label{a6}
\ee
where $E'(r)=\sum_{\mu\ne\nu}(\gamma^{\mu}\gamma^{\nu}\frac{1}{2}
[S_{\mu}\,,\,S_{\nu}\,]+ir\gamma^{\mu}[S_{\mu}\,,\,C_{\nu}\,])$ has a bound
$||E'(r)||\le{}c'(r)\,\e$. A simple generalisation of the argument in 
\cite{Neu(bound)} shows that this bound is satisfied with 
$c'(r)=(1+r\sqrt{2})d(d-1)/2$. Following \cite{Kerler,Neu(bound)} we also note that
for an eigenvalue $\lambda(m,r)=\la\psi(m,r)\,,\,H(m,r)\psi(m,r)\ra$ we have
$\frac{d}{dm}\lambda(m,r)=-r\la\psi(m,r)\,,\,\g5\psi(m,r)\ra$ and consequently
$|\frac{d}{dm}\lambda(m,r)|\le{}r$, which implies $|H(m,r)|\,\ge\,|H(m',r)|-r|m'-m|$.

We consider the cases $r\le1$ and $r\ge1$ separately. In the former case
(\ref{a6}) together with (\ref{3.10}) gives
\be
H(m,r)^2&=&r^2\Big(\,\sum_{\mu}S_{\mu}^2+(-m+\sum_{\mu}R_{\mu})^2\Big)
+(1-r^2)\sum_{\mu}S_{\mu}^2+E'(r) \nonumber \\
&\ge&r^2(1+\chi(m))-c'(r)\,\e
\label{a7}
\ee
This together with (\ref{3.15}) gives $H(k,r)^2\ge{}r^2(1-(c_k''+c'(r)/r^2)\,\e)$,
and consequently, setting $c_k(r)=c_k''+c'(r)/r^2\,$, 
\be
|H(m,r)\;\ge\;r\sqrt{1-c_k(r)\,\e}-r|k-m|\quad,\qquad\ k=1,3,\dots,2d-1
\label{a8}
\ee
In the $r\ge1$ case we rewrite (\ref{a6}) as 
\be
H(m,r)^2&=&\sum_{\mu}S_{\mu}^2+(-m+\sum_{\mu}R_{\mu})^2
+(r^2-1)(-m+\sum_{\mu}R_{\mu})^2+E'(r) \nonumber \\
&\ge&1+\chi(m)-c'(r)\,\e
\label{a9}
\ee
and it follows from (\ref{3.15}) that 
\be
|H(m,r)|\;\ge\;\sqrt{1-\tilde{c}_k(r)\,\e}-r|k-m|\quad,\qquad\ k=1,3,\dots,2d-1
\label{a10}
\ee
with $\tilde{c}_k(r)=c_k''+c'(r)$.

Note that the bounds (\ref{a8}) for the $r<1$ case and (\ref{a10}) for the 
$r>1$ case are both weaker than the bound (\ref{2.2}) for the $r\!=\!1$ case.

\end{document}